\pdfoutput=1
\documentclass[sigconf]{acmart}

\AtBeginDocument{%
  }

\copyrightyear{2026}
\acmYear{2026}
\setcopyright{cc}
\setcctype{by}
\acmConference[WWW '26]{Proceedings of the ACM Web Conference 2026}{April 13--17, 2026}{Dubai, United Arab Emirates}
\acmBooktitle{Proceedings of the ACM Web Conference 2026 (WWW '26), April 13--17, 2026, Dubai, United Arab Emirates}
\acmPrice{}
\acmDOI{10.1145/3774904.3792790}
\acmISBN{979-8-4007-2307-0/2026/04}

\usepackage{multirow}
\usepackage{enumitem}
\usepackage{balance}

\begin{document}

\title{Beyond the Flat Sequence: Hierarchical and Preference-Aware Generative Recommendations}

\settopmatter{authorsperrow=4}

\author{Zerui Chen}
\orcid{0009-0009-3226-619X}
\authornote{Both authors contributed equally to this research.}
\email{zrchen@ir.hit.edu.cn}
\affiliation{%
  \institution{Harbin Institute of Technology}
  \city{Harbin}
  \country{China}
}

\author{Heng Chang}
\orcid{0000-0002-4978-8041}
\authornotemark[1]
\email{changh.heng@gmail.com}
\affiliation{
  \institution{Huawei Technologies Co., Ltd.}
  \city{Beijing}
  \country{China}
}

\author{Tianying Liu}
\email{tianying_liu@outlook.com}
\affiliation{
  \institution{Huawei Technologies Co., Ltd.}
  \city{Shanghai}
  \country{China}
}

\author{Chuantian Zhou}
\email{alita@bupt.edu.cn}
\affiliation{
  \institution{Beijing University of Posts and Telecommunications}
  \city{Beijing}
  \country{China}
}

\author{Yi Cao}
\email{caoyi23@huawei.com}
\affiliation{
  \institution{Huawei Technologies Co., Ltd.}
  \city{Shanghai}
  \country{China}
}

\author{Jiandong Ding}
\email{dingjiandong2@huawei.com}
\affiliation{
  \institution{Huawei Technologies Co., Ltd.}
  \city{Shanghai}
  \country{China}
}

\author{Ming Liu}
\authornote{Corresponding author.}
\email{mliu@ir.hit.edu.cn}
\affiliation{
  \institution{Harbin Institute of Technology}
  \city{Harbin}
  \country{China}
}

\author{Bing Qin}
\email{qinb@ir.hit.edu.cn}
\affiliation{
  \institution{Harbin Institute of Technology}
  \city{Harbin}
  \country{China}
}

\renewcommand{\shortauthors}{Zerui Chen et al.}

\begin{abstract}
  Generative Recommenders (GRs), exemplified by the Hierarchical Sequential Transduction Unit (HSTU), have emerged as a powerful paradigm for modeling long user interaction sequences. However, we observe that their "flat-sequence" assumption overlooks the rich, intrinsic structure of user behavior. This leads to two key limitations: a failure to capture the temporal hierarchy of session-based engagement, and computational inefficiency, as dense attention introduces significant noise that obscures true preference signals within semantically sparse histories, which deteriorates the quality of the learned representations. To this end, we propose a novel framework named \textbf{HPGR} (\textbf{H}ierarchical and \textbf{P}reference-aware \textbf{G}enerative \textbf{R}ecommender), built upon a two-stage paradigm that injects these crucial structural priors into the model to handle the drawback. Specifically, HPGR comprises two synergistic stages. First, a structure-aware pre-training stage employs a session-based Masked Item Modeling (MIM) objective to learn a hierarchically-informed and semantically rich item representation space. Second, a preference-aware fine-tuning stage leverages these powerful representations to implement a Preference-Guided Sparse Attention mechanism, which dynamically constrains computation to only the most relevant historical items, enhancing both efficiency and signal-to-noise ratio. Empirical experiments on a large-scale proprietary industrial dataset from APPGallery and an online A/B test verify that HPGR achieves state-of-the-art performance over multiple strong baselines, including HSTU and MTGR.
\end{abstract}

\ccsdesc[500]{Information systems~Recommender systems}

\keywords{attention mechanism, generative recommendation}

\maketitle

\begin{figure}[htbp]
  \centering
  \includegraphics[width=0.9\linewidth]{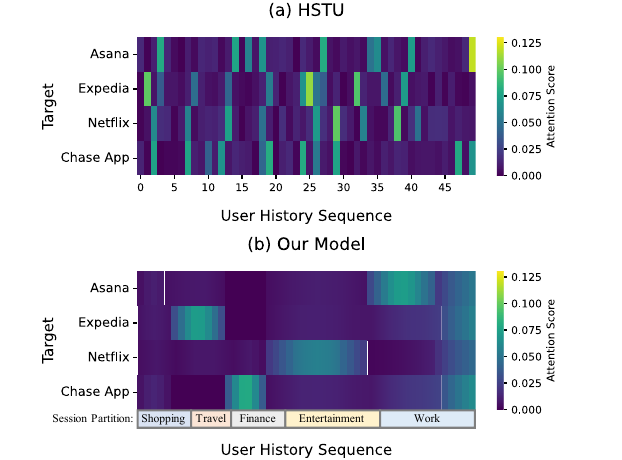}
  \caption{A comparison of attention mechanisms visualizing how different models interpret user history. (a) The baseline HSTU model, which treats history as a "flat-sequence", exhibits scattered, item-level attention that fails to capture the underlying structure of user intent. (b) In contrast, HPGR learns to understand the hierarchical, session-based nature of user behavior.}
  \label{fig:opening_graph}
\end{figure}

\section{Introduction}
Recommendation systems form the cornerstone of modern internet platforms, personalizing experiences for billions of users daily.

Recently, the field has witnessed a paradigm shift from traditional Deep Learning Recommendation Models (DLRMs) \cite{zhang2019deep, guo2017deepfm, covington2016deep, cheng2016wide, zhou2018deep, tang2020progressive, wang2021dcn} towards Generative Recommenders (GRs) \cite{zhai2024actions}. Spearheading this movement, the Hierarchical Sequential Transduction Unit (HSTU) \cite{zhai2024actions} has achieved monumental success, scaling recommendation models to trillions of parameters and demonstrating promising scaling laws akin to those of Large Language Models (LLMs) \cite{minaee2024large, achiam2023gpt, touvron2023llama, bai2023qwen}.

By reformulating recommendation as a sequential transduction task over long user histories, HSTU effectively overcomes the scalability bottlenecks of DLRMs and unifies the modeling of heterogeneous features into a single, powerful  architecture.

Despite its impressive capabilities in processing long sequences, the success of HSTU is predicated on a fundamental yet implicit assumption: it models a user's history as a flat, undifferentiated stream of actions. Although effective, the "flat-sequence" perspective overlooks the rich intrinsic structure inherent in human behavior, leading to two significant challenges.

First, it neglects hierarchical temporal structure of user engagement. User activities are not uniformly distributed but are naturally clustered into sessions, each representing a focused, short-term intent. A flat model struggles to distinguish coherent intra-session patterns from broader inter-session interest shifts \cite{yang2024item}. 

Second, it overlooks the sparse nature of user preference, whereby the next action depends on a compact, highly relevant subset of past interactions instead of the full historical record. Compelling the model to compute dense attention across thousands of historical items is not only computationally redundant but also introduces significant noise that can impair the quality of learned representations \cite{chaudhari2021attentive}.

To address these limitations, we propose HPGR, a new framework that injects crucial structural priors into the generative recommendation process. Our core insight is that by explicitly modeling the temporal hierarchy and preference of user behavior, we can build recommenders that are both more effective and more efficient. 

Specifically, we introduce a two-stage paradigm. In the first stage, called structure-aware pre-training, we introduce a Session Enhancement Module (SEM) and employ a session-based Masked Item Modeling (MIM) objective. The SEM explicitly models the temporal hierarchy of user behavior by processing interactions within and across sessions. This process compels the model to learn generalizable item semantics and complex behavioral dynamics, resulting in a powerful pre-trained embedding space.

In the second stage termed preference-aware fine-tuning, we adapt the pre-trained model for the downstream prediction task. It is achieved through our novel Preference-Guided Sparse Attention (PGSA) mechanism, which aims to reconstruct the most salient preference context for a given candidate item. Rather than applying dense attention to the entire history, PGSA treats the candidate as a query to retrieve a small, highly relevant subset of past interactions. Attention is then restricted to this preference-rich subset, sharply reducing computation and improving accuracy by amplifying genuine preference signals while suppressing historical noise.

Our work makes the following key contributions:
\begin{itemize}[topsep=0pt, itemsep=-2pt, parsep=0pt]
    \item We propose the Session Enhancement Module (SEM), a hierarchical transformer architecture, and a corresponding structure-aware pre-training strategy. This combined approach enables the model to explicitly capture both intra- and inter-session dynamics, overcoming the critical "flat-sequence" limitation of prior generative recommenders.
    \item We introduce the Preference-Guided Sparse Attention (PGSA) mechanism for the fine-tuning stage. PGSA replaces prior random sampling techniques with a preference-driven sparsity. This approach improves the accuracy of attention by focusing on the most relevant signals, while being inherently more computationally efficient than dense attention for long user sequences.
    \item We conduct extensive offline experiments and an online A/B test on a large-scale industrial recommender system. The results validate that HPGR substantially outperforms state-of-the-art baselines and achieves a significant +1.99\% eCPM uplift in a real-world production environment, demonstrating its practical effectiveness and business value.
\end{itemize}

\begin{figure*}[t]
  \centering
  \includegraphics[width=0.9\linewidth]{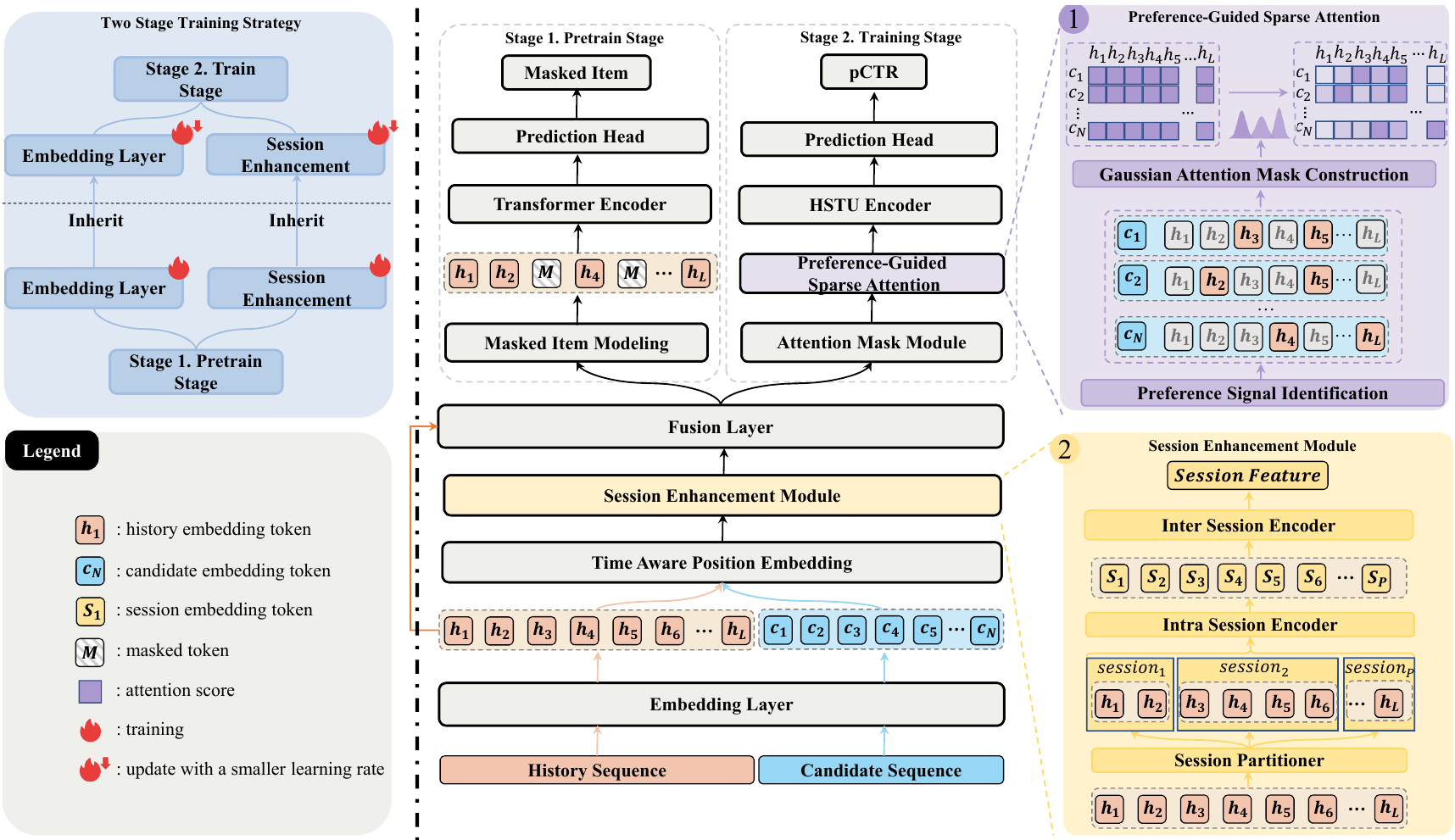}
  \caption{The overall architecture of our proposed HPGR framework. The model operates in a two-stage training paradigm. The central flow illustrates the shared foundation, including the Session Enhancement Module (SEM), which provides hierarchical context. The architecture then branches into two distinct stages: (1) Stage 1 (Pre-training), which uses a Transformer Encoder and a Masked Item Modeling objective to learn structurally-aware item representations. (2) Stage 2 (Fine-tuning), which employs an HSTU Encoder for the downstream pCTR task, enhanced by our novel Preference-Guided Sparse Attention (PGSA) mechanism. The panels on the right provide detailed views of the hierarchical structure of the SEM and the operational steps of PGSA. The top-left panel highlights the weight inheritance strategy, where pre-trained modules are updated with a smaller learning rate during fine-tuning.}
  \label{fig:framework_graph}
\end{figure*}

\section{Related Work}
\subsection{Sequential and Generative Recommenders}
Traditional sequential recommenders aim to capture user dynamics from interaction sequences, evolving from RNN-based models like GRU4Rec \cite{hidasi2015session} to attention-based architectures like SASRec \cite{kang2018self}. In large-scale industrial systems, these approaches are often realized as target attention modules within a larger DLRM framework.

Recently, a paradigm shift towards generative recommenders has emerged, inspired by the success of Large Language Models \cite{minaee2024large, achiam2023gpt, touvron2023llama, bai2023qwen}. Instead of treating sequences as just one of many feature sources, GRs reformulate the entire recommendation task as a sequence-to-sequence problem over a unified stream of user actions. The seminal work on HSTU \cite{zhai2024actions} demonstrated that this approach can scale to trillions of parameters and exhibits promising scaling laws. Follow-up works like MTGR \cite{han2025mtgr} have further adapted this architecture to better incorporate the rich cross-features traditionally used in DLRMs. Our work builds directly upon this powerful GR paradigm. However, while existing GRs treat the user history as a "flat-sequence", we argue for the importance of incorporating explicit structural priors, a key distinction that our HPGR framework addresses.

\subsection{Modeling Structure in User Sequences}
The concept that user behavior is not monolithic but organized into sessions of intent is well-established in recommendation literature, particularly in session-based recommendation \cite{wang2021survey, hidasi2018recurrent}. These works typically focus on predicting the next action within a short, ongoing session. However, in the context of modeling very long user histories, as championed by the GR paradigm, explicitly modeling the hierarchical structure of multiple, interconnected sessions is a less explored area \cite{ma2019hierarchical, li2019multi}. Most long-sequence models either treat the sequence as flat \cite{kang2018self, sun2019bert4rec} or use implicit mechanisms like positional biases \cite{vaswani2017attention} to distinguish recent from older items. Our work addresses this gap by introducing the Session Enhancement Module (SEM), a hierarchical transformer architecture designed to explicitly model both intra- and inter-session dynamics within long user sequences, providing a richer, structurally-aware context for prediction.

\subsection{Efficient Attention for Long Sequences}
The quadratic complexity of self-attention remains a significant bottleneck for long-sequence modeling \cite{zhuang2023survey, vaswani2017attention}. A vast body of research has focused on improving attention efficiency through methods like low-rank approximations \cite{wang2020linformer} or hardware-aware implementations like FlashAttention \cite{dao2022flashattention, dao2023flashattention, shah2024flashattention}. Another line of work introduces sparsity into the attention matrix \cite{child2019generating, beltagy2020longformer}. The original HSTU model, for instance, proposes a form of stochastic length reduction inspired by stochastic depth \cite{huang2016deep, zhai2024actions}.

While effective, these methods are typically content-agnostic or rely on random sampling. In contrast, our PGSA offers a more principled, content-aware approach. It is inspired by the observation that for a given candidate item, only a small subset of the user's history is truly relevant. By leveraging the learned representations to dynamically identify this preference-rich subset for each prediction, our method introduces a powerful inductive bias, creating a form of informed sparsity that is guided by the task itself, differing fundamentally from prior random or fixed sparsity patterns.

\subsection{Pre-training for Recommendation}
The pre-train, fine-tune paradigm, popularized by models like BERT \cite{devlin2019bert} and GPT \cite{achiam2023gpt}, has also gained traction in the recommendation domain \cite{zeng2021knowledge, qiu2021u, liu2023pre}. One prominent trend involves leveraging pre-trained LLMs to enhance recommendation through their rich world knowledge and in-context learning capabilities \cite{wu2024survey, geng2022recommendation}. Another trend focuses on pre-training user representations from their behavior sequences, which can then be transferred to various downstream tasks \cite{zhou2020s3}.

Our work aligns with this latter trend but proposes a novel pre-training objective tailored to our goals. Instead of learning a generic user representation, our structure-aware pre-training stage utilizes our novel SEM to explicitly model the behavioral hierarchy. This architecture is then trained with a session-based masked item modeling objective \cite{sun2019bert4rec} to simultaneously co-learn rich item semantics and the dynamics of the user's hierarchical behavior. This creates a powerful foundation that is explicitly prepared for our downstream model, which relies on both semantic understanding and hierarchical context.

\section{Method}
Our proposed framework, HPGR, enhances the generative recommendation paradigm by explicitly modeling the inherent structure of user behavior. It comprises a two-stage training framework: (1) a structure-aware pre-training stage to learn general-purpose, hierarchically-informed item representations, and (2) a preference-aware fine-tuning stage to adapt the model to specific downstream tasks through an efficient Preference-Guided Sparse Attention (PGSA) mechanism, which leverages the pre-trained knowledge to focus on the most relevant preference signals in the user's history. Figure~\ref{fig:framework_graph} illustrates the overall architecture of our framework. 

\subsection{Stage 1: Structure-aware Pre-training}
The primary goal of the pre-training stage is twofold: first, to learn a high-quality item embedding space that captures rich semantic relatedness, and second, to train a powerful Session Enhancement Module (SEM) capable of understanding the hierarchical nature of user interests. Both components are learned in a self-supervised manner, independent of any specific downstream task, to form a robust foundation for fine-tuning.

\subsubsection{Session Enhancement Module}
To move beyond the "flat-sequence" assumption, we introduce a SEM that models user history hierarchically. Given a user's long interaction history $H = \{h_1, h_2, \dots, h_N\}$, we first partition it into a sequence of sessions $S = \{S_1, S_2, \dots, S_M\}$ based on temporal inactivity. The module then employs a two-level Transformer \cite{vaswani2017attention} architecture:
\begin{itemize}
    \item \textbf{Intra-session transformers}: A shared Transformer encoder is applied to each session $S_i$ independently, modeling the local, short-term dependencies among items within a coherent interaction context. The output representation of the \texttt{[CLS]} token is then taken as the summary vector $s_i$ for that session.
    \item \textbf{Inter-session transformer}: The sequence of session summary vectors $[s_1, s_2, \dots, s_M]$ is fed into a second, higher-level transformer encoder, which models the long-term evolution and transitions of user interests across different sessions.
\end{itemize}

Formally, for each session $S_i$, the intra-session transformer produces a summary vector $s_i$:
\begin{equation}
s_i = \text{Intra-Transformer}(\text{Emb}(S_i))_{\texttt{[CLS]}}
\end{equation}
where $\text{Emb}(S_i)$ denotes the embedded item sequence of session $S_i$, and $(\cdot)_{\texttt{[CLS]}}$ indicates selecting the output representation of the \texttt{[CLS]} token. Subsequently, the inter-session transformer processes the sequence of all session vectors to yield the final global session context vector, $F_{\text{sess}}$:
\begin{equation}
F_{\text{sess}} = \text{Inter-Transformer}([s_1, s_2, \dots, s_M])_{\text{GlobalPool}}
\end{equation}
where $(\cdot)_{\text{GlobalPool}}$ could be another \texttt{[CLS]} token's output or a mean pooling operation over all session representations. The context vector $F_{\text{sess}}$ is then integrated into the item representations by broadcast-wise addition to the initial token embeddings of the entire history sequence: $H_{\text{enhanced\_emb}} = H_{\text{emb}} + F_{\text{sess}}$.

\subsubsection{Pre-training objective}
We employ a masked item modeling objective for pre-training \cite{sun2019bert4rec}. For a given user, let $H_m$ be the set of masked item indices in their session-enhanced history $H_{\text{enhanced}}$. The pre-training objective is to minimize the negative log-likelihood of predicting the correct masked items, formulated as:
\begin{equation}
\mathcal{L}_{\text{MIM}} = - \sum_{i \in H_m} \log P(h_i | H_{\text{unmasked}})
\label{eq:mim_loss}
\end{equation}
where $P(h_i | H_{\text{unmasked}})$ is the probability of predicting the original item $h_i$ given the unmasked context, computed via a softmax layer over the output of a transformer encoder. This task compels the model to learn deep semantic relationships between items and user behavior patterns, resulting in a robust and meaningful item embedding table. The outputs of this stage are the pre-trained item embeddings $\mathbf{E}_{\text{pre}}$ and the weights of the SEM $\mathbf{\Theta}_{\text{SEM}}$.

\subsection{Stage 2: Preference-aware Fine-tuning}
In the PAF stage, we adapt the model for the downstream pCTR prediction task. The core idea of this stage is to inherit the pre-trained weights from Stage 1: the item embedding layer and the SEM are initialized with $\mathbf{E}_{\text{pre}}$ and $\mathbf{\Theta}_{\text{SEM}}$ respectively. The architecture is then enhanced with the following components for fine-tuning.

\begin{figure}[t!]
  \centering
  \includegraphics[width=0.7\linewidth]{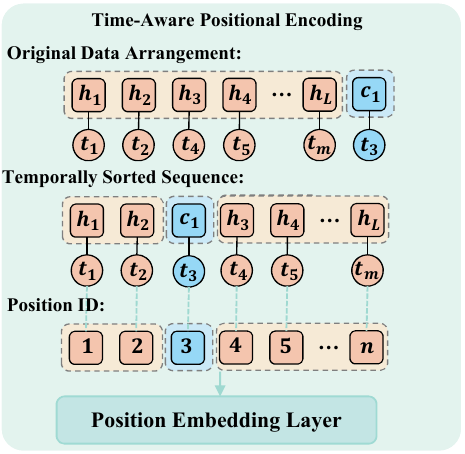}
  \caption{The workflow of Time-Aware Positional Encoding. First, the candidate item $c_1$ is inserted into the user's history sequence at its correct chronological position, forming a complete temporally sorted sequence. Then, a new set of positional IDs is assigned to this final sequence, ensuring that each item's position reflects its precise temporal order before being passed to the embedding layer.}
  \label{fig:timeaware_graph}
\end{figure}

\subsubsection{Time-Aware Positional Encoding}
To precisely model the temporal relationship between historical items and the candidate item $C$, we replace standard positional encoding with a time-aware strategy that constructs a chronologically consistent set of indices. As illustrated in Figure~\ref{fig:timeaware_graph}, the process involves two steps: First, given the history sequence $H = \{h_1, \dots, h_N\}$ and a candidate item $C$, we form a new, unified sequence by inserting $C$ into its correct position based on its timestamp. Second, we assign a contiguous set of positional indices $\{1, 2, \dots, N+1\}$ to this temporally sorted sequence. The specific positional index for the candidate item $C$ is formally determined by:
\begin{equation}
\text{Index}(C) = 1 + \sum_{i=1}^{N} \mathbb{I}(t_i < t_c)
\end{equation}
where $t_i$ is the timestamp of the historical item $h_i$, and $\mathbb{I}(\cdot)$ is the indicator function which equals 1 if the condition is true, and 0 otherwise. This ensures all items in the final sequence have a unique and temporally correct positional representation.

\subsubsection{Preference-Guided Sparse Attention}
To overcome the inefficiency of applying dense attention over long histories with thousands of items, we propose Preference-Guided Sparse Attention (PGSA). PGSA constructs a compact yet powerful preference context for each candidate item, and concentrates modeling capacity on the most salient signals of user preference. In particular, this is accomplished in three steps:

\textbf{(i) Preference signal identification}: For a given candidate item $C$, we first identify potential preference signals within the user's history. Concretely, we compute the preference alignment between $C$ and every historical item $h_i$ through the dot product of their pre-trained embeddings: $\text{score}_i = \text{emb}(C) \cdot \text{emb}(h_i)$. A higher score indicates stronger preference alignment.

\textbf{(ii) Preference subset selection}: We then select the Top-K historical items with the highest preference alignment scores. The selected items with indices $\{p_1, p_2, \dots, p_K\}$ form a sparse subset that acts as the primary \textbf{anchors of user preference} for the candidate, moving beyond mere semantic similarity.

\textbf{(iii) Gaussian attention mask construction}: Rather than using a hard binary mask, we construct a soft attention mask that applies a gaussian-shaped influence around each preference anchor. The final mask value $M_j$ for any position $j$ in the history is calculated as Equation~(\ref{eq:gaussian_mask}):
\begin{equation}\label{eq:gaussian_mask}
M_j = \max_{k \in \{1, \dots, K\}} \left\{ \exp\left( - \frac{(j - p_k)^2}{2\sigma^2} \right) \right\}
\end{equation}
where $\sigma$ is a hyperparameter controlling the width of the influence. This mask is applied element-wise to the attention score matrix within the HSTU encoder, which concentrates computation on regions of high preference alignment, maintaining smooth gradients while effectively filtering out noise from irrelevant interactions.

The session-enhanced token embeddings, together with their time-aware positional embeddings and the gaussian mask, are fed into the main HSTU encoder. The resulting candidate representation is then passed through a multi-layer perceptron to produce the predicted score.

\subsubsection{Fine-tuning Objective}
The entire HPGR model is trained end-to-end in this stage to predict the click-through rate. Given a training batch of user-candidate pairs, the objective is to minimize the binary cross-entropy loss, defined as:
\begin{equation}
\mathcal{L}_{\text{pCTR}} = - \frac{1}{B} \sum_{i=1}^{B} \left( y_i \log(\hat{y}_i) + (1 - y_i) \log(1 - \hat{y}_i) \right)
\label{eq:pctr_loss}
\end{equation}
where $B$ is the batch size, $y_i \in \{0, 1\}$ is the ground-truth label (click or no-click), and $\hat{y}_i$ is the predicted CTR score for the $i$-th pair, produced by the final MLP and a sigmoid activation function.

\subsection{Training Strategy}
Our two-stage approach ensures both general knowledge acquisition and task-specific adaptation. 
The process begins with the structure-aware pre-training stage, where the model is trained on a large corpus of unlabeled user interaction data to optimize $\mathcal{L}_{\text{MIM}}$ (Eq.~\ref{eq:mim_loss}). 

Subsequently, in the preference-aware fine-tuning stage, we initialize the core modules with the learned weights and train the entire HPGR model on the labeled downstream task by optimizing $\mathcal{L}_{\text{pCTR}}$ (Eq.~\ref{eq:pctr_loss}). During this second stage, we apply a smaller learning rate to the inherited weights (item embeddings and SEM) compared to the rest of the randomly-initialized model parts. This differential learning rate strategy is crucial: it preserves the rich, general-purpose semantic space learned during pre-training while allowing the model to effectively adapt to the specific supervised signals of the downstream task.

\section{Experiments}
\subsection{Experimental Setup}

\subsubsection{Dataset}
To validate our approach in a real-world, large-scale industrial setting, we constructed a proprietary dataset from the production recommender system of the \textbf{APPGallery} application distribution platform. The full dataset captures one continuous month of user-interaction logs. Table~\ref{tab:dataset_stats} summarizes the statistics of this \textbf{complete, unsampled dataset}, highlighting its massive scale.

To facilitate extensive experimentation, including hyperparameter tuning and ablation studies, our main experiments (reported in Table~\ref{tab:main_results}) were conducted on a representative subset. This subset was created by \textbf{randomly sampling 2.5\% of the users}, along with their complete interaction histories. All data was rigorously anonymized to protect user privacy.

To confirm that the conclusions drawn from the sampled data generalize to the full data distribution, we also conducted a final validation on the complete test set, as detailed in Section~\ref{sec:full_dataset_validation}.

\begin{table}[htbp]
\centering
\caption{Statistics of the APPGallery Dataset.}
\label{tab:dataset_stats}
\begin{tabular}{lr}
\toprule
\textbf{Metric} & \textbf{Value} \\
\midrule
Number of Users       & $\sim$45.8 Million \\
Number of Items       & $\sim$200,000 \\
Interactions (Train) & $\sim$285 Million \\
Interactions (Test)  & $\sim$7.38 Million  \\
\bottomrule
\end{tabular}
\end{table}

\subsubsection{Baselines}
To comprehensively evaluate our proposed model, HPGR, we compare it against a wide range of baseline methods, which we categorize based on their underlying technical paradigm.

\begin{itemize}
    \item \textbf{Discriminative Recommendation Models:} This group includes a spectrum of models that formulate recommendation as a scoring or classification task. It ranges from foundational architectures to the current state-of-the-art in high-order feature interaction modeling.
    \begin{itemize}
        \item \textbf{DNN \cite{covington2016deep}}: A fundamental deep learning model that processes features with multi-layer perceptrons.
        \item \textbf{MoE \cite{ma2018modeling}}: An extension of DNN that uses a mixture-of-experts architecture to handle diverse user populations.
        \item \textbf{GRU4Rec \cite{hidasi2015session}}: A classic model that uses gated recurrent units to capture temporal patterns in user behavior sequences.
        \item \textbf{SASRec \cite{kang2018self}}: A seminal work that first introduced self-attention to effectively capture long-range dependencies in user sequences.
        \item \textbf{BERT4Rec \cite{sun2019bert4rec}}: A model inspired by BERT that uses a deep bidirectional self-attention network for sequential recommendation.
        \item \textbf{Wukong \cite{zhang2024wukong}}: A state-of-the-art model that establishes a scaling law for recommendation via a "dense scaling" approach, using stacked Factorization Machines to capture high-order feature interactions.
    \end{itemize}

    \item \textbf{Generative Recommendation Models:} This category represents the emerging paradigm that our work builds upon. These models reformulate recommendation as a sequence generation task, showing great promise in modeling long user histories.
    \begin{itemize}
        \item \textbf{HSTU \cite{zhai2024actions}}: The pioneering work on generative recommendations and the direct foundation of our research.
        \item \textbf{MTGR \cite{han2025mtgr}}: A powerful successor in the GR paradigm, which further enhances the architecture for handling long sequences in industrial systems.
    \end{itemize}
\end{itemize}

For our proposed model, \textbf{HPGR}, we experiment with several configurations of different scales to validate its scalability, as detailed in our subsequent analyses.

\begin{table*}[htbp]
  \centering
  \caption{Performance comparison on our industrial dataset. The best result is in \textbf{bold}, and the second best is \underline{underlined}. Impr.\% represents the relative improvement of the best HPGR model compared to the strongest baselines (underlined).}
  \label{tab:main_results}
  \begin{tabular}{llr}
    \toprule 				
    \textbf{Category} & \textbf{Model} & \textbf{AUC ($\uparrow$)} \\
    \midrule
    \multirow{6}{*}{\textbf{Discriminative Models}} 
    & DNN (\citeauthor{covington2016deep} \citeyear{covington2016deep}) & 0.7923 \\
    & MoE (\citeauthor{ma2018modeling} \citeyear{ma2018modeling}) & 0.8024 \\
    & GRU4Rec (\citeauthor{hidasi2015session} \citeyear{hidasi2015session}) & 0.8146 \\
    & BERT4Rec (\citeauthor{sun2019bert4rec} \citeyear{sun2019bert4rec}) & 0.8145 \\
    & SASRec (\citeauthor{kang2018self} \citeyear{kang2018self}) & 0.8194 \\
    & Wukong (\citeauthor{zhang2024wukong} \citeyear{zhang2024wukong}) & 0.8167 \\
    \midrule

    \multirow{2}{*}{\textbf{Generative Models}} 
    & HSTU (\citeauthor{zhai2024actions} \citeyear{zhai2024actions}) & 0.8220 \\
    & MTGR (\citeauthor{han2025mtgr} \citeyear{han2025mtgr}) & \underline{0.8253} \\
    \midrule

    \multirow{4}{*}{\textbf{Our Model}} 
    & HPGR (Base, Pre-training only) & 0.8327 \\
    & \quad + PGSA module & 0.8348 \\
    & \quad + SEM module & 0.8374 \\
    & \textbf{HPGR (Full)} & \textbf{0.8377} \\
    \midrule

    \multicolumn{2}{l}{\textbf{Impr.}} & \textbf{+1.50\%} \\
    \bottomrule
  \end{tabular}
\end{table*}

\subsection{Main Results}
To evaluate the effectiveness of our proposed framework HPGR, we conducted a comprehensive comparison against all baseline models on the APPGallery dataset. The main results for the ranking task, grouped by their underlying paradigm, are presented in Table~\ref{tab:main_results}. From the table, we can draw several key observations:

\begin{enumerate}
    \item \textbf{Performance within the Discriminative Paradigm:} The first group of models, representing the discriminative paradigm, shows a clear trend where explicitly modeling user sequences is beneficial. Sequential models like SASRec and GRU4Rec significantly outperform their non-sequential counterparts such as DNN. Notably, SASRec, with its self-attention mechanism, emerges as the strongest model within this entire paradigm, setting the performance ceiling for discriminative approaches on this dataset.

    \item \textbf{Superiority of the Generative Paradigm:} A crucial finding is that the generative paradigm demonstrates a fundamental performance advantage. Even the baseline generative models, HSTU and MTGR, surpass the best-performing discriminative model, SASRec. This validates that reformulating recommendation as a sequence generation task is a more powerful approach for modeling long and complex user histories in large-scale industrial settings.

    \item \textbf{HPGR Sets a New State-of-the-Art:} Our proposed model, \textbf{HPGR}, not only affirms the strength of the generative approach but also significantly pushes its boundary. The full HPGR model achieves the best performance overall with an AUC of \textbf{0.8377}. This represents a substantial uplift over the strongest baseline, MTGR, with a +0.0124 absolute AUC gain and a \textbf{+1.50\%} relative improvement. This result strongly demonstrates that explicitly injecting structural priors is a critical and highly effective enhancement to the generative recommendation framework.
\end{enumerate}

\subsection{Validation on the Full Dataset}
\label{sec:full_dataset_validation}
\begin{figure}[htbp]
  \centering
  \includegraphics[width=0.7\linewidth]{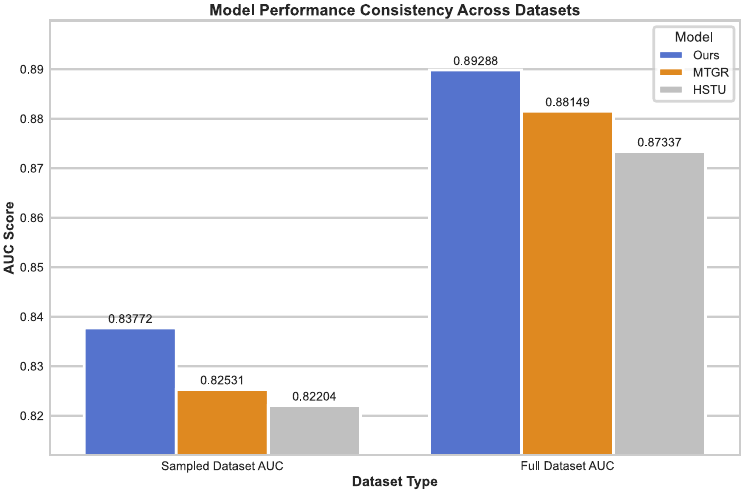}
  \caption{Performance consistency of the top models across the sampled and full test datasets.}
  \label{fig:full_dataset_validation}
\end{figure}

To confirm that the superior performance observed on the sampled subset is robust and generalizable, we conducted a  validation experiment on the complete, unsampled test dataset. 
As shown in Figure~\ref{fig:full_dataset_validation}, the relative performance ranking of the top models remains consistent. 
On this full dataset, HPGR achieves an impressive AUC of \textbf{0.89288}, maintaining a significant lead of \textbf{+0.01139} absolute AUC points over the strongest baseline, MTGR. 
This result provides strong evidence that the advantages conferred by our architectural innovations are fundamental and translate effectively to the complete, real-world data distribution, confirming the generalizability of our main findings.

\subsection{Ablation Study}
To dissect the individual contributions of our key innovations—the structure-aware pre-training, the PGSA mechanism, and the SEM architecture—we conducted a detailed ablation study. The results are presented in the lower section of Table~\ref{tab:main_results}, starting from a base configuration and progressively adding components.

\begin{itemize}
    \item \textbf{Impact of Pre-training:} The \textbf{HPGR (Base, Pre-training only)} model utilizes the standard HSTU architecture but initializes its item embeddings from our structure-aware pre-training stage (using the MIM objective). It achieves an AUC of 0.8327. This result, which already outperforms the strongest baseline MTGR, powerfully validates that our pre-training objective alone produces a fundamentally superior and more generalizable item embedding space compared to task-specific end-to-end training.

    \item \textbf{Impact of PGSA:} Adding the PGSA mechanism to the base model (\textbf{+ PGSA module}) improves the AUC from 0.8327 to 0.8348. This gain demonstrates that dynamically focusing attention on a sparse, preference-rich subset of historical items is highly effective at reducing noise and amplifying relevant signals, especially for long sequences.

    \item \textbf{Impact of SEM:} When we instead add only the SEM to the base model (\textbf{+ SEM module}), the AUC rises substantially to 0.8374. This significant improvement underscores the critical importance of explicitly modeling the temporal hierarchy of user behavior, which is overlooked by the flat-sequence assumption of the base architecture.

    \item \textbf{Synergy of All Components:} Finally, the \textbf{HPGR (Full)} model, which integrates both SEM and PGSA, yields the best overall result. This demonstrates the synergy of our core innovations, highlighting the value of both capturing the hierarchical structure of user behavior and focusing attention on relevant historical items within the generative paradigm, leading to state-of-the-art performance.
\end{itemize}

\subsection{Qualitative Analysis of the Pre-trained Embedding Space}

To validate our structure-aware pre-training, we analyze the item embedding table before and after the MIM stage. Figure~\ref{fig:embedding_analysis} compares Xavier initialization ("Random Init") with our pre-trained embeddings ("Train"). This comparison reveals three key structural transformations:

\begin{figure}[htbp]
  \centering
  \includegraphics[width=0.5\textwidth]{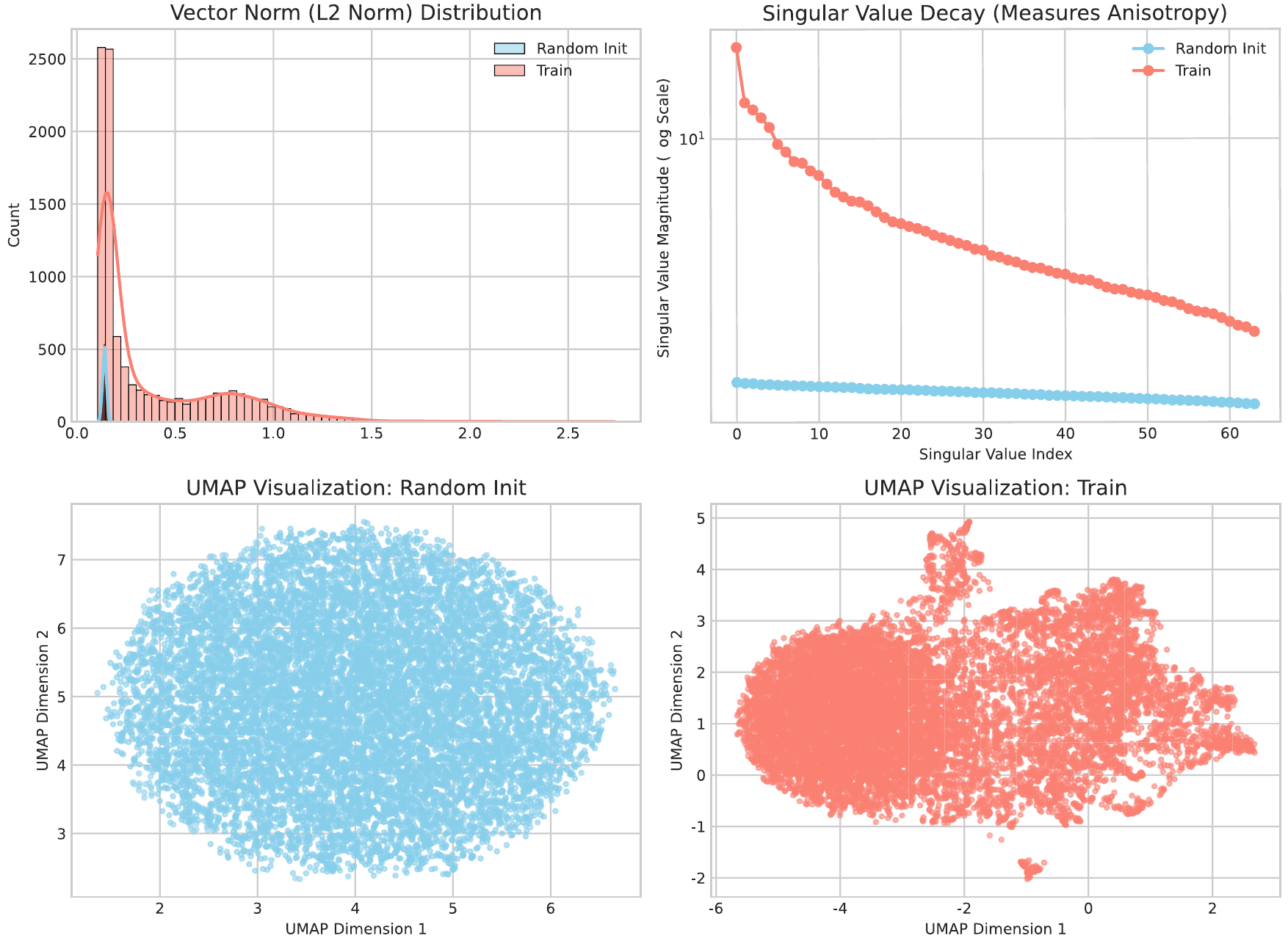}
  \caption{Comparison of embedding properties between Random Initialization and Pre-training. Top left: L2 norm distribution shifts to a high-density peak. Bottom row: UMAP visualization reveals that the embeddings organize into meaningful semantic patterns, in contrast to the initial random cloud.}
  \label{fig:embedding_analysis}
\end{figure}

\begin{itemize}[leftmargin=*, nosep]
    \item \textbf{Learned magnitude shift}: The L2 norm distribution (top left) shifts from uniform to highly skewed. Most items, likely representing the long-tail, collapse into a low-norm region. Conversely, a subset of semantically significant items differentiates with higher norms, indicating the model learns to compress less prominent items while highlighting important ones.

    \item \textbf{Emergence of anisotropy}: The singular value decay (top right) indicates a transition from an isotropic to an anisotropic structure. This confirms the model utilizes specific dimensions to capture principal axes of semantic variance rather than spanning the space uniformly.

    \item \textbf{Structured Semantic Landscape}: UMAP visualizations (bottom row) show a transformation from an unstructured cloud to distinct substructures. The pre-trained embeddings organize into high-density regions, demonstrating that semantically similar items are successfully pulled into proximity.
\end{itemize}

\subsection{Hyperparameter Study} \label{sec:hyper_study}
\begin{figure}[htbp]
  \centering
  \includegraphics[width=0.7\linewidth]{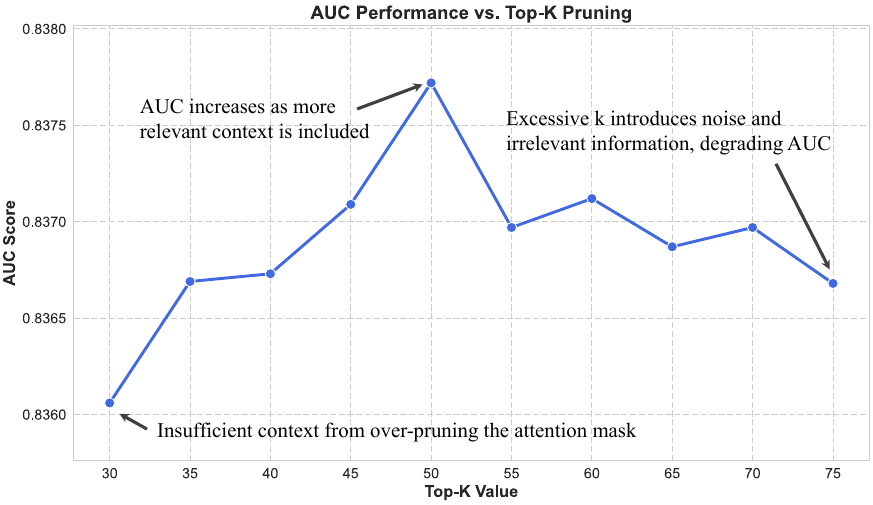}
  \caption{Hyperparameter analysis of AUC with respect to the Top-K value.}
  \label{fig:topk_sensitivity}
\end{figure}

We investigated the sensitivity of the Top-K value, $K$, in our PGSA mechanism. As shown in Figure~\ref{fig:topk_sensitivity}, performance peaks at $K=50$, indicating  that captures sufficient context without excessive noise. We use $K=50$ for all other experiments.

\subsection{Scalability Analysis}
\begin{figure}[htbp]
  \centering

  \includegraphics[width=0.7\linewidth]{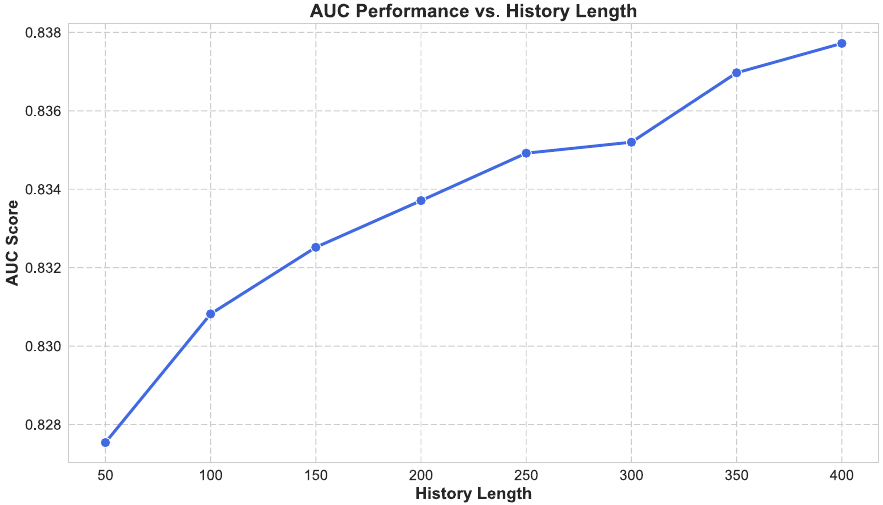}

  \includegraphics[width=0.7\linewidth]{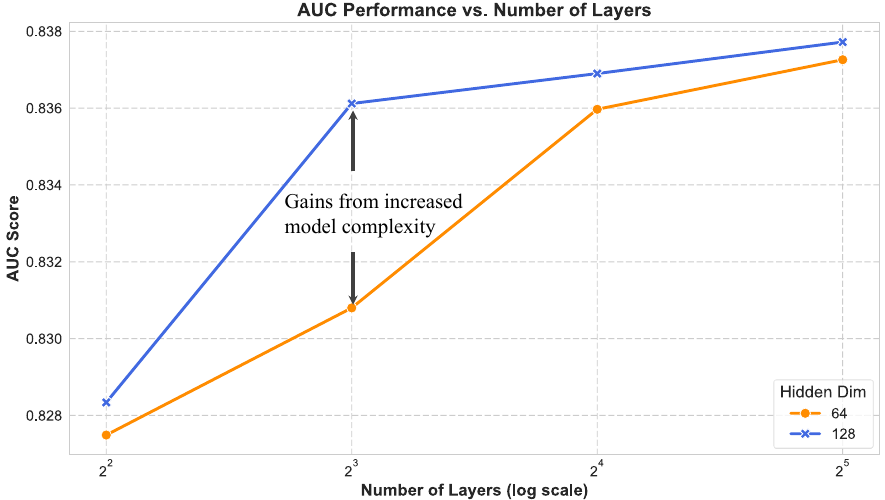}

  \caption{Scalability of HPGR with respect to history length (top) and model size (bottom).}
  \label{fig:scaling_plots}
\end{figure}

We analyzed the scaling properties of HPGR. As shown in Figure~\ref{fig:scaling_plots}, performance consistently improves with longer user histories and with increased model depth and width. This demonstrates that HPGR is a robust and scalable architecture, promising further gains with more data and computational resources.

\subsection{Efficiency Analysis}
\begin{figure}[htbp]
  \centering
  \includegraphics[width=0.7\linewidth]{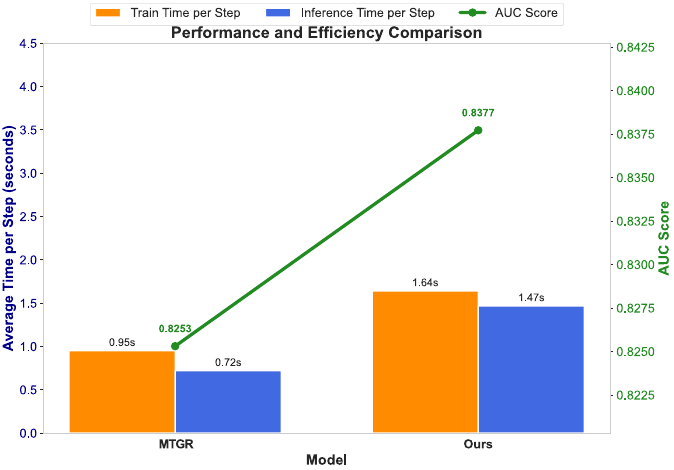}
  \caption{Comparison of training time, inference time, and AUC score between MTGR and our HPGR model.}
  \label{fig:efficiency_comparison}
\end{figure}

We also compared the computational efficiency of HPGR against MTGR. Figure~\ref{fig:efficiency_comparison} shows that while our model incurs a manageable increase in training and inference time, this investment yields a substantial gain of +0.0124 in absolute AUC. This highly favorable trade-off demonstrates the practical value of our architectural enhancements.

\section{Online Evaluation}

To ultimately validate the real-world impact and practical value of our proposed HPGR framework, we deployed our model for an online A/B test on the \textbf{APPGallery} App recommendation platform. The experiment was conducted over a 7-day period, where traffic was split evenly between our proposed HPGR model and the incumbent, a highly optimized production DLRM, which served as the online baseline.

We focused our evaluation on the primary business indicator, \textbf{eCPM (Effective Cost Per Mille)}, which directly measures platform revenue and is a critical metric for industrial recommender systems. The aggregated results, reported as the relative improvement over the online baseline, are summarized in Table~\ref{tab:online_results}.

\begin{table}[htbp]
  \centering
  \caption{Online A/B test results, with offline AUC evaluated on the full dataset. This comparison shows that HPGR's offline superiority translates to a significant online eCPM uplift.}
  \label{tab:online_results}
  \begin{tabular}{c|cc} 
    \toprule
    \textbf{Model} & \shortstack{\textbf{Offline AUC}} & \shortstack{\textbf{Online eCPM}} \\
    \midrule
    Production DLRM (Baseline) & 0.8880          & Baseline \\ 
    HPGR (ours)                & \textbf{0.8929} & \textbf{+1.99\%} \\ 
    \bottomrule
  \end{tabular}
\end{table}
The online A/B test results robustly confirm the effectiveness of our model in an online production environment. As shown in the table, HPGR achieved a statistically significant +1.99\% uplift in eCPM over the highly optimized baseline. 

\section{Conclusion}
In this paper, we introduce HPGR, a novel framework that addresses the "flat-sequence" limitation in modern Generative Recommenders. HPGR injects crucial structural priors via a two-stage paradigm: a structure-aware pre-training stage with a Session Enhancement Module (SEM) to learn hierarchical representations, followed by a preference-aware fine-tuning stage using Preference-Guided Sparse Attention (PGSA) to focus on relevant history. This dual approach enables the model to capture complex, session-based user behavior. Extensive offline experiments and an online A/B test validate HPGR's effectiveness, demonstrating state-of-the-art performance and a significant +1.99\% eCPM uplift in a production environment. 

\begin{acks}
The research in this article is supported by the National Science Foundation of China (U22B2059, 62276083), Key Research and Development Program of Heilongjiang Province (2024ZX01A05) and the 5G Application Innovation Joint Research Institute’s Project (A003).
\end{acks}

\bibliographystyle{ACM-Reference-Format}
\balance
\bibliography{base}

\end{document}